\documentclass[amssymb,twocolumn,aps]{revtex4}
\usepackage{graphics,color,graphicx,amsmath}
\usepackage{subcaption}
\usepackage{natbib}
\usepackage{verbatim}
\usepackage{graphicx}
\usepackage[above,below]{placeins}
\usepackage{float}
\usepackage{tabularx}
\usepackage{times}
\usepackage{blindtext}
\usepackage{url}
\usepackage{rotating}
\usepackage{mathtools}
\usepackage{threeparttable}
\usepackage{enumitem}

\setlist{nosep}                 

\begin{document}

  \title{Measuring fidelity of implementation of named active learning methods in physics}

  \author{Ibukunoluwa Bukola~\footnote[1]{ib396@drexel.edu}$^{1}$, Meagan Sundstrom$^{1}$, Justin Gambrell$^{2}$, Colin Green$^{3}$, Adrienne L. Traxler$^{4}$, and Eric Brewe~\footnote[2]{Contact author: eb573@drexel.edu}$^1$}
    \affiliation{$^{1}$Department of Physics, Drexel University, Philadelphia, Pennsylvania 19104, USA\\
    $^{2}$Department of Computational Mathematics, Science and Engineering, Michigan State University, East Lansing, Michigan 48824, USA\\
    $^3$Department of Physics, Bryn Mawr College, Bryn Mawr, Pennsylvania 19010, USA\\
    $^{4}$Department of Science Education, University of Copenhagen, Copenhagen, Denmark}


  \begin{abstract}
Various active learning methods have been developed for introductory physics, and these methods are increasingly being adopted by instructors. However, instructors often do not implement these methods exactly as was originally intended by the developers, as they may face issues related to funding and institutional support for active learning and/or have different instructional contexts (e.g., student populations) and environments (e.g., physical classroom layouts) than the developers. Existing research does not sufficiently capture the range of variation in instructor implementation of established active learning methods, especially in comparison to high-fidelity implementations. In this study, we first identify the critical components (i.e., components without which the active learning method cannot be said to have been implemented) of three named active learning methods: SCALE-UP, ISLE, and Tutorials. We then evaluate the fidelity with which 18 different introductory physics instructors implement these methods by analyzing classroom observations and comparing the extent to which these broader implementations use each critical component in their classroom to high-fidelity implementations. We find across all three active learning methods that broader implementations spend similar amounts of class time on the critical components as high-fidelity implementations. At the same time, we observe substantial variation in the specific styles that broader implementers operationalize these critical components (e.g., doing a few long activities versus many short activities). Finally, we find no clear relationship between fidelity of implementation and student conceptual learning gains for our study's sample of instructors, providing preliminary evidence that different ways of implementing the critical components of active learning method may all effectively improve student understanding.
  \end{abstract}

  \maketitle

\section{Introduction}

The benefits of active learning to student outcomes, such as conceptual learning~\cite{freeman2014active, terenzini2001collaborative} and persistence~\cite{braxton2008role, miller2021supporting}, are well-documented in the literature. Research also shows that over the past decade, the number of college-level physics instructors using research-based instructional strategies, including named active learning methods (e.g., Peer Instruction~\cite{mazur1997peer}), has significantly increased~\cite{dancy2024physics}. As these named methods become more widely implemented, instructors are modifying the methods based on their local contexts, needs, and/or constraints~\cite{affriyenni2025navigating}. Instructors, for example, may not receive appropriate funding~\cite{gess2003educational} and/or institutional support~\cite{foote2014diffusion,henderson2007barriers,affriyenni2025navigating} to facilitate their use of active learning. They may also have different student populations, class sizes, and physical classrooms than the original developers~\cite{foote2014diffusion, henderson2007barriers}.

Research has demonstrated that this flexibility of active learning methods to individual instructor modifications may help increase adoption and lead to improvements of the methods~\cite{rogers2003diffusion}. However, some modifications may depart too far from the established best practices~\cite{willison2023examining}. Instructors, for example, may unknowingly omit \textit{critical components} of the method, or components without which the method cannot be said to have been implemented~\cite{dancy2016faculty}. Consequently, many existing studies on the effectiveness of research-based instructional strategies are met with concerns about the \textit{fidelity} with which the instructors in the studies implemented the active learning methods they use, i.e., the extent to which instructors implemented the method in the way that the original developers intended~\cite{hake1998interactive, mowbray2003fidelity, foote2014diffusion}. Indeed, one study found that active learning does not improve student conceptual understanding more than traditional lecturing among a random set of instructors; rather, active learning may be more effective when implemented by science education researchers~\cite{andrews2011active}. 
Together, these concerns and existing studies point to a growing need to better understand how instructors implement active learning methods and the effects of fidelity of implementation on student outcomes, such as conceptual learning.


Several previous studies have measured fidelity of implementation in undergraduate science courses~\cite{dancy2016faculty,borrego2013fidelity, scanlon2019method}. Borrego and colleagues~\cite{borrego2013fidelity}, for example, collected surveys from 387 engineering science faculty who used different research-based teaching practices. The authors measured fidelity as whether or not each instructor spent class time on each critical component of the research-based practices they implemented. The researchers found a wide range in fidelity across the examined active learning methods, spanning 11\% to 80\% of instructors implementing all critical components, and that methods with only one critical component exhibited higher fidelity than methods with multiple critical components. 

The above studies, however, rely on survey responses from or interviews with instructors which are prone to bias, hence, may not accurately represent what instructors are actually doing in their classrooms~\cite{dancy2016faculty}. Furthermore, to our knowledge, there are no studies that directly relate fidelity of implementation to student learning in the context of undergraduate science. The current study, therefore, uses direct classroom observations to characterize fidelity, leveraging the Classroom Observation Protocol for Undergraduate STEM (COPUS)~\cite{smith2013classroom} to accurately measure instructional practices. We measure the fidelity with which 18 introductory physics instructors (hereon referred to as ``broader" implementations") implement three named active learning methods: Student Centered Activities for Large Enrollment Undergraduate Programs (SCALE-UP) or Studio Physics, Investigative Science Learning Environment (ISLE), and Tutorials. We quantify fidelity in two ways: descriptively, comparing the fractions of class time spent on each activity in the COPUS to those of a high-fidelity implementation (including a focus on the COPUS activities that directly reflect each method's critical components; we define high-fidelity implementations to be those at the site where the method was developed or based on the recommendation of experts, as in Ref.~\cite{commeford2021characterizing}), and with network analysis, comparing the temporal transitions between COPUS activities to those of a high-fidelity implementation~\cite{Sundstrom2025}. We also directly relate these measures of fidelity to student understanding using concept inventory data. 
We aim to address the following research questions: 
\begin{enumerate}
    \item How does instructor use of critical components vary between high-fidelity and broader implementations of named active learning methods, as measured by relevant COPUS activities? 
     \item To what extent does a broad set of instructors implement named active-learning methods with fidelity, based on full COPUS observations?
    \item How, if at all, is fidelity of implementation related to student conceptual learning?
\end{enumerate}


\section{Background}

In this section, we summarize existing literature about fidelity of implementation, critical components, and our analytic methods.


\subsection{Fidelity of implementation}

Fidelity of implementation has been extensively studied and defined in various fields, but only more recently in the Discipline-Based Education Research (DBER) community~\cite{stains2017fidelity}. Broadly, fidelity of implementation is defined as a measure of closeness of an implemented intervention to the original intervention~\cite{o2008defining}. In the DBER context, Stains and colleagues define fidelity of implementation as ``the extent to which the critical components of an intended educational program, curriculum, or instructional practice are present when that program, curriculum, or practice is enacted"~\cite{stains2017fidelity} (p. 2). We operationalize this definition in the current study, where we consider active learning methods to be the relevant intervention.

\begin{figure}[t]
    \includegraphics[width=0.48\textwidth]{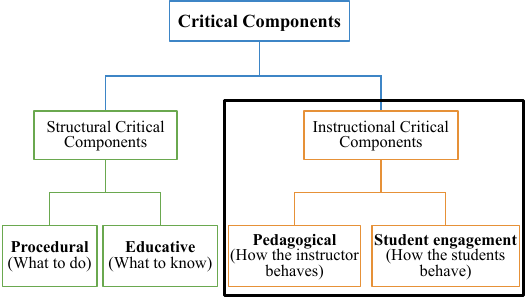}
    \caption{Types of critical components as defined by Stains and colleagues~\cite{stains2017fidelity}. The outlined black box highlights the focus of this study: instructional critical components.}\label{criticalcomponent}
\end{figure}

As Stains and colleagues mention, identifying the critical components of an intervention (i.e., components without which the intervention cannot have been said be carried out) is central to fidelity studies~\cite{century2010framework, mowbray2003fidelity, borrego2013fidelity}. 
There are two types of critical components: structural and process (or instructional, as in Ref.~\cite{stains2017fidelity}; Fig.~\ref{criticalcomponent}). Structural critical components are related to program adherence (e.g., material covered, duration of intervention, and what knowledge the implementer must possess)~\cite{harn2013balancing,borrego2013fidelity}. Process or instructional critical components are related to how the intervention is implemented (e.g., instructor and student behaviors)~\cite{mowbray2003fidelity,century2010framework}. In this study, we focus on the instructional critical components, as some studies have noted that these components more directly impact student outcomes~\cite{gersten2005quality,mowbray2003fidelity}.  We determine the instructional critical components of each active learning method through extensive literature review, similar to Ref.~\cite{borrego2013fidelity} (see further details in Methods).

Fidelity studies also require measurement and validation of fidelity of implementation. Fidelity can be measured using researcher or expert ratings from observations or documentation, user surveys, or interviews~\cite{borrego2013fidelity, stains2017fidelity}. Researcher and expert ratings, however, are subjective and may be prone to biases, and user surveys and interviews may be inaccurate. In this study, we measure fidelity by applying a structured observation protocol to direct classroom observations and mapping the critical components to corresponding  codes in the protocol (see further details in the next section and in Methods). We validate fidelity by comparing the observations of broader implementations to those of high-fidelity implementations of the active learning methods. 

\begin{table*}[t]
  \caption{Summary of COPUS codes for instructor and student activities. Full definitions can be found in Ref.~\cite{smith2013classroom}. \label{copuscodes}}
  \begin{ruledtabular}
    \begin{tabular}{ll}
      Code & Definition \\ 
      \hline
      Instructor\\
      \hspace{0.5cm} Lec & Lecturing about course material to the whole class\\
     \hspace{0.5cm}  RtW & Real-time writing on the board \\
     \hspace{0.5cm}  FUp & Following up on clicker question or activity to whole class \\
     \hspace{0.5cm}  PQ & Posing question to the whole class\\
     \hspace{0.5cm}  CQ & Asking a clicker question to the class \\
     \hspace{0.5cm}  AnQ\_I & Answering questions from students with the entire class listening \\
     \hspace{0.5cm}  MG & Moving and guiding ongoing student work during active learning task \\
    \hspace{0.5cm}   1o1 & One-on-one extended discussion with one or more individuals \\
     \hspace{0.5cm}  D/V & Demonstration, video, experiment, simulation, or animation \\
     \hspace{0.5cm}  Adm & Administrative tasks \\
    \hspace{0.5cm}   W & Waiting \\
    \hspace{0.5cm}   O & Other \\
    Student \\
    \hspace{0.5cm}   L & Listening to instructor \\
    \hspace{0.5cm}   Ind & Individual thinking or problem solving \\
     \hspace{0.5cm}  CG & Answering clicker question in groups \\
    \hspace{0.5cm}   WG & Working in groups on worksheet \\
     \hspace{0.5cm}  OG & Other group activity (e.g., laboratory experiment) \\
     \hspace{0.5cm}  AnQ\_S & Answering a question posed by the instructor \\
    \hspace{0.5cm}   SQ & Student asks question to instructor in front of the whole class \\
    \hspace{0.5cm}   WC & Whole-class discussion \\
     \hspace{0.5cm}  Prd & Predicting the outcome of demonstration or experiment \\
    \hspace{0.5cm}   SP & Student presentation to the whole class\\
     \hspace{0.5cm}  TQ & Test or quiz \\
    \hspace{0.5cm}   W & Waiting \\
    \hspace{0.5cm}   O & Other \\
    \end{tabular}
  \end{ruledtabular}
\end{table*}

\subsection{Classroom Observation Protocol for Undergraduate STEM (COPUS)}

We measure fidelity by characterizing the instructor and student activities taking place during class using direct observation. Various classroom observation protocols have been developed with this aim~\cite{anwar2021systematic}. These protocols are typically either open-ended or structured. Open-ended protocols expect observers to provide answers to certain prompts based on their observations, while structured protocols require observers to check off the presence of a specified list of activities. We aim to compare observations of different instructors; therefore, we opted to use a structured protocol for our study. We use the COPUS~\cite{smith2013classroom}, a structured observation protocol with 25 specified codes to characterize both instructor (12 codes) and student (13 codes) behaviors in the classroom (Table~\ref{copuscodes}). COPUS observers check off codes that are present in two-minute time intervals. 

We use the COPUS in the current study because this protocol includes both instructor and student codes and we aim to characterize both of these facets of instructional critical components (Fig.~\ref{criticalcomponent}). Stains and colleagues, moreover, identify the COPUS as the only tool that ``aligns with the measure of some of the potential critical components"~\cite{stains2017fidelity} (p. 9) of educational interventions. Additionally, in this study, we use data from a combination of live and video-recorded classroom observations. COPUS is well suited for both types of observations and affords the ability to measure inter-coder reliability among multiple coders.  

COPUS was designed to facilitate descriptive studies about instructional practices~\cite{smith2013classroom}, and many existing research studies use descriptive statistics to analyze COPUS data (e.g., calculating the fraction of two-minute time intervals that each code is present)~\cite{smith2014campus,commeford2021characterizing}. Instructional practices, however, are more complex than descriptive measures may capture, as they involve both instructor and student behaviors that occur dynamically over time. To gain a more nuanced understanding of teaching practices, researchers have increasingly turned to more advanced analytical approaches, such as clustering, mixture modeling, and network analysis of COPUS data~\cite{commeford2022characterizing,stains2018anatomy,sundstrom2025relativebenefits,Sundstrom2025}. 

We use both descriptive measures and network analysis in the current study so that our measures of fidelity capture a rich representation of instructional practices. These two conceptualizations of fidelity are distinct, but complementary: descriptive measures tell us the extent to which certain activities are implemented, while network analysis tells us about the specific style in which these activities are implemented.

\subsection{Network analysis}

Network analysis is a method for analyzing complex systems~\cite{grunspan2014,brewe2018roles}. Networks consists of nodes, which are the actors or entities in the system of interest, and edges, which indicate connections or relationships between the nodes. In DBER, network analysis has been used in multiple ways, though its primary application is in \textit{social} network studies, where the researchers examine relationships among people (e.g., students). Commeford and colleagues, for example, used network analysis to examine how patterns of peer interactions change over the course of a semester in different active learning methods~\cite{commeford2021characterizing}. Other studies have used network analysis to study student reasoning when solving physics problems~\cite{bodin2012mapping, speirs2024inferential}. 

In this study, we apply network analysis to our COPUS observations to capture the complex and temporal nature of instruction, similar to Ref.~\cite{Sundstrom2025}. We represent classroom activities (i.e., the COPUS codes) as nodes in our observation networks and the chronological transitions between activities as the edges. Here we aim to compare broad implementations of active learning methods to high-fidelity implementations, and treating classroom observations as networks allows us to quantify fidelity by calculating a single similarity measure between pairs of classroom observations (see further details in Methods).




\section{Methods}

\begin{table*}[t]
  \caption{Critical components of the three examined active learning methods and corresponding instructor and student COPUS codes (see definitions in Table~\ref{copuscodes}).\label{criticalcomponents-table}}
  \begin{ruledtabular}
    \begin{tabular}{llll}
 Method& Critical component& Instructor codes&Student codes \\ 
 \hline
SCALE-UP~\cite{beichner2007student,beichner2000introduction, beichner1999student, beichner2008scale} & Restaurant-style classrooms &&\\
& Lab, lecture, and recitation integrated together &&\\
& New material introduced through pre-class readings and quizzes &&\\
& Students work on tangibles, ponderables, and experiments in groups & MG & OG\\
& Class-wide discussions after completing activities & FUp, PQ & AnQ\_S\\
& Students present their answers &  & SP\\

ISLE~\cite{etkina2007investigative,etkina2021investigative,etkina2006using,tufino2025exploring} & TA training&&\\
&Students complete readings outside of class&&\\
& Students devise explanations for observation experiments in groups & MG, D/V & OG\\
& Students create and predict outcomes of testing experiments in groups & MG & OG, Prd\\
& Students solve problems with classroom response system & CQ, MG & CG, Ind\\
Tutorials~\cite{mcdermott2002tutorials,finkelstein2005replicating, docktor2014synthesis} & TA training&&\\
& Students complete pre-test and post-test &  & \\
& Students complete worksheets in groups & MG & WG\\

    \end{tabular}
  \end{ruledtabular}
\end{table*}

We examine fidelity of implementation of three well-established and widely implemented active learning methods used to teach introductory physics and astronomy: SCALE-UP, ISLE, and Tutorials. In this section, we describe the critical components of each of these methods as well as our data sources and analysis methods. De-identified data and analysis scripts can be found at Ref.~\cite{github2025}.

\subsection{Identifying critical components of active learning methods}

Using the fidelity of implementation framework (Fig.~\ref{criticalcomponent})~\cite{mowbray2003fidelity,century2010framework,stains2017fidelity}, we reviewed existing literature to determine the critical components of each method (Table~\ref{criticalcomponents-table}). This literature included the foundational papers describing each method and research studies examining various aspects of these methods. Importantly, the authors of one of these research studies had their descriptions of the active learning methods checked by the method developers, increasing the validity of their descriptions~\cite{commeford2021characterizing}. We considered a course element to be a critical component if the developers noted the element as part of the curriculum and/or if the aforementioned study described the element in their expert-validated description. The first two authors engaged in negotiated discussions until consensus was reached about the final set of critical components for each method, which are described in the subsections below.

We then identified all COPUS codes that would indicate the presence of each critical component by mapping the descriptions of the critical components onto the definitions of the COPUS activities (some critical components did not have corresponding COPUS codes; Table~\ref{criticalcomponents-table}). For example, if laboratory experiments were a critical component, we expected the instructor to be moving around and guiding group work (MG code in the COPUS) while the students work on the group work (OG code in the COPUS because the group work here is experimental rather than in the form of clicker questions or worksheets). Similar to identifying the critical components, the first two authors iteratively discussed the relevant COPUS codes until consensus was reached.

\subsubsection{SCALE-UP}

SCALE-UP was developed to ``scale up" the Studio Physics pedagogy, which was originally intended for small classrooms~\cite{beichner2007student,beichner2000introduction, beichner1999student, beichner2008scale}. One of the critical components of the SCALE-UP method is the physical classroom layout: the classrooms are often described as ``restaurant-style," where each table holds two to three groups of three to four students each. All course components (i.e., lectures, labs, and recitations) are integrated into one classroom meeting time, which often occurs three times per week for two hours. In-class time is focused on group work, where students work on ``tangibles" (e.g., lab activities) and ``ponderables" (e.g., conceptual, open-ended questions that encourage critical thinking, estimations, and problem solving)~\cite{beichner2007student}. Instructors serve as coaches during class time, circling through the classroom and helping students come up with answers to their own questions. After completing activities, students are encouraged to present their answers to the class, and their answers are then reviewed and discussed by peers and instructors. New material is typically introduced through pre-class readings and assessed through quizzes completed before the class meetings.

\subsubsection{ISLE}

ISLE can be implemented in all components of a class, or it can be implemented only within the lab component~\cite{etkina2007investigative,etkina2021investigative,etkina2006using,tufino2025exploring}. Critical to this method is that students observe an experiment and provide multiple explanations (i.e., scientific hypotheses) for the observations and then conduct or design an experiment to test their explanations. Based on the results of their testing experiment, they either reject, revise, or accept their explanations. This cycle allows students to engage in the scientific process as they learn. Students also engage in quantitative problem solving in ISLE classes, typically using elements of Peer Instruction~\cite{mazur1997peer}, such as classroom response systems (e.g., clickers). Students are assigned readings after in-class sessions to emphasize that scientific knowledge is sourced from scientific processes. Teaching assistants for ISLE classes usually receive training prior to class sessions. Worksheets that incorporate most of these aspects of ISLE have been developed and popularized for use in introductory physics labs~\cite{ISLEweb}. 

\subsubsection{Tutorials}

Tutorials for Introductory Physics~\cite{mcdermott2002tutorials,finkelstein2005replicating, docktor2014synthesis} and Astronomy~\cite{adams2003lecture} are typically implemented in either the recitation or lecture component of a course. Tutorials consist of students working on worksheets in groups, with the teaching assistant or instructor moving through groups to address student questions. The Tutorial worksheets are carefully designed to elicit and confront common student misconceptions. 

Meetings with teaching assistants are usually held before the tutorials are implemented to familiarize them with the worksheets. Pre-tests are usually administered after relevant topics are covered during lecture, but before the Tutorials are implemented. After the Tutorials, a post-test is administered to students. 

\subsection{Data sources}

This study uses data from two iterations of a national research project titled ``Characterizing Active Learning Environments in Physics" (CALEP): classroom observations of high-fidelity implementations of each method (from the first iteration of the project, CALEP1), and classroom observations and concept inventory data from 18 adopters of the three methods (from the second iteration of the project, CALEP2).

\subsubsection{CALEP1: High-fidelity implementations}

In the first iteration of the project, researchers collected classroom observation data from one high-fidelity implementation of several active learning methods (Table~\ref{instructor-info})~\cite{commeford2021characterizing}. The classroom observations captured all class meetings within one week and were conducted live and in-person by a single observer using the COPUS. The high-fidelity implementation sites were chosen as either the institution where the active learning method was developed, or an institution recommended by instructors with extensive research experience on the methods. Further details of the data collection procedures for the CALEP1 study can be found in Ref.~\cite{commeford2021characterizing}. In the current study, we analyze the COPUS data from the SCALE-UP, ISLE, and Tutorials courses collected in this prior work. 


\subsubsection{CALEP2: Broader implementations}

As part of the current iteration of the project~\cite{Sundstrom2025,sundstrom2025relativebenefits}, we recruited 18 introductory physics and astronomy instructors across the United States who self-reported using SCALE-UP, ISLE, or Tutorials in their introductory physics or astronomy course. 
Instructors were recruited through the research team's personal networks, advertisements on the American Physical Society website, and/or identification through grant advisory board members, as we sought to collect data from a broad range of implementations. All courses took place during the fall 2023, spring 2024, or fall 2024 semesters. 

Seven SCALE-UP instructors, four ISLE instructors, and seven Tutorials instructors participated in the study (Table~\ref{instructor-info} and Table~\ref{average-classsize} in the Appendix). One Tutorials course was astronomy, and all other courses were physics. Within the Tutorials courses, there were two types of implementation: whole-class implementations (four courses), where the Tutorials worksheets were implemented as part of lecture sections, and recitation-only implementations (three courses), where the Tutorials worksheets were only implemented in recitation sections. We analyzed these different implementations separately, as noted in the Results. 

\begin{table}[t]
  \caption{Types of institutions represented in the two data sources. For CALEP1, instructors came from three unique institutions in the United States (one per method)~\cite{commeford2021characterizing}. For CALEP2, each of the 18 instructors came from a different institution in the United States. Carnegie Classifications of Research Activity are from 2025: R1 indicates ``Very High Research Spending and Doctorate Production," R2 indicates ``High Research Spending and Doctorate Production," and RCU indicates ``Research Colleges and Universities."    \label{instructor-info}}
  \begin{ruledtabular}
    \begin{tabular}{lcc}
 Type of Institution & CALEP1 & CALEP2\\

  \hline
 Public/Private& 3/0&11/7\\
 Highest Degree Awarded&\\
 \hspace{3mm} Associate's & 0& 1\\
 \hspace{3mm} Bachelor's & 0&2\\
 \hspace{3mm} Master's &0& 6\\
 \hspace{3mm} Ph.D. &3& 9\\
 Carnegie Classification&\\
\hspace{3mm} R1 &3& 5\\
\hspace{3mm} R2 & 0&3\\
\hspace{3mm} RCU & 0& 4\\
Hispanic-Serving Institution & 0 & 3\ \\
 \end{tabular}
  \end{ruledtabular}
\end{table}

\begin{table*}[t]
  \caption{Summary of the data used in our study by active learning method: the subset of observations from the CALEP1 COPUS data included in our analysis, the number of courses with three video observations in the CALEP2 data, and the number of CALEP2 courses with concept inventory data.\label{observations}}
  \begin{ruledtabular}
    \begin{tabular}{lccc}
 Method& CALEP1 classroom & CALEP2 courses with & CALEP2 courses with \\
 &observations&three video observations & concept inventory data\\
  \hline
SCALE-UP & 3 class sessions & 7&5\\
ISLE & 1 lecture and 1 recitation &  4&3\\
Tutorials (Recitation-only) & 1 recitation &  3&3\\
Tutorials (Whole-class) & 1 lecture and 1 recitation & 4& 3\\
    \end{tabular}
  \end{ruledtabular}
\end{table*}

We collected two types of data from each instructor: video recordings of class sessions and pre- and post-semester concept inventory scores. Instructors recorded class sessions using their own camera, Zoom, or a camera we sent to them by mail. They recorded three consecutive classroom meetings to capture a typical week in their course, as recommended in prior literature~\cite{stains2018anatomy}. All 18 instructors recorded three full class sessions and are included in our fidelity analysis (i.e., the first and second research questions). 

Instructors also collected concept inventory data through research-validated assessments administered online at the beginning and end of the semester. Instructors chose a concept inventory that best aligned with their course content. Most instructors used the Force Concept Inventory~\cite{hestenes1992force} or Force and Motion Concept Evaluation~\cite{ramlo2008validity} (see Table~\ref{conceptinventory-breakdown} in the Appendix). Out of the 18 instructors, 14 had at least 50\% of enrolled students who took both the pre- and post-concept inventory (i.e., had matched responses) and are included in our analysis of student conceptual learning (i.e., the third research question). Each instructor received \$1,000 as compensation for participating in the study.

\subsection{Data analysis}

\subsubsection{COPUS coding}

As mentioned, we used the COPUS to capture student and instructor behaviors during class. The CALEP1 observations were conducted live as part of a prior study~\cite{commeford2021characterizing}, so we used the existing COPUS coding of those courses. These observations, however, included a mix of course components (i.e., lectures and recitations) and, in some cases, several different sections of students conducting the same activity (e.g., several recitation sections of a large-enrollment course using Tutorials). To ensure that the CALEP1 and CALEP2 data were comparable for our fidelity analysis, we used a subset of the CALEP1 observations with course structures that most aligned with those of the observations collected for CALEP2 in our analysis (Table~\ref{observations}). In cases where the CALEP1 data included observations of different sets of students completing the same activity (i.e., recitation sections of both ISLE and Tutorials), we used a random number generator to select one observation to include in the current analysis (no CALEP2 instructors recorded multiple sections of students doing the same activity). For Tutorials, the recitation section selected from the CALEP1 data was used as the high-fidelity comparison for the recitation-only implementations in CALEP2. We aggregated this same section's observation with a lecture section observation when comparing CALEP1 to the CALEP2 whole-class implementations of Tutorials to increase comparability (the CALEP2 whole-class implementations embedded Tutorials worksheets within their lecture sections). The CALEP1 data also included two different implementations of ISLE: lab only and whole-class (i.e., ISLE implemented in lecture, recitations, and labs). We only used the whole-class implementation data from CALEP1, as its structure was more similar to that of the ISLE courses in CALEP2. 

The second, third, and fourth authors applied the full COPUS to the 54 classroom video recordings collected in CALEP2 (i.e., three video recordings for each of the 18 instructors). First, all three authors independently coded one video per active learning method. Then, the authors iteratively met to discuss disagreements and individually re-coded the videos until an inter-relater reliability of over 80\% was reached for each method (as measured with Cohen's Kappa~\cite{landis1977measurement}; see values in Ref.~\cite{sundstrom2025relativebenefits}). After reaching this level of agreement, we randomly assigned one of these three authors to code each video in its entirety.

We included all 25 COPUS activities in our coding, however we excluded three instructor codes (1o1, O, and W; Table~\ref{copuscodes}) and one student code (WC, Table~\ref{copuscodes}) from the current analysis because these codes were rarely marked as being present (i.e., each of these codes were observed in less than 2\% of all two-minute time intervals in the dataset). We also aggregated the observations for each instructor to get a more holistic representation of their instructional style. For the CALEP1 data, we aggregated lecture and recitation observations for ISLE and the whole-class implementation of Tutorials. For the CALEP2 data, we combined each instructor's three video recordings (all CALEP2 instructors recorded the same course component for all three observations).

\subsubsection{Constructing classroom observation networks}

For each set of observations considered in our comparisons (i.e., four from CALEP1 and 18 from CALEP2, Table~\ref{observations}), we constructed two classroom observation networks: one for instructor activities and one for student activities. The network nodes were the COPUS codes. The network edges were directed, with arrows indicating a chronological transition from one code to another (i.e., pointing from one code to the code that happened afterward). As in Ref.~\cite{Sundstrom2025}, we considered a transition between codes to be when a code was not present in a two-minute time interval and then became present in the following two-minute time-interval. We created an edge from all codes in the previous time interval to the newly appearing code(s) in the following interval. The network edges were also weighted as the number of occurrences of each transition divided by the total number of observed two-minute time intervals (i.e., to normalize for different class durations). Though not directly part of the network structure, we also calculated the fraction of class time spent on each COPUS code as the fraction of the observed two-minute time intervals that the code was present. We represented these fractions of class time in the network diagrams in the node colors (Fig.~\ref{network-example}).


\subsubsection{Measuring fidelity}

To address our first research question, we compared the fraction of class time spent on each COPUS code relevant to the identified critical components between the high-fidelity (CALEP1) and broader (CALEP2) implementations (Table~\ref{criticalcomponents-table}). We conducted this analysis descriptively by comparing the high-fidelity implementation measurement to the distribution of broader implementations measurements for each active learning method.

To address our second research question, we calculated four similarity metrics for each pair of high-fidelity and broader implementation within the same active learning method: two that compare the fraction of class time spent on each COPUS code (one measure for instructor codes and one measure for student codes, including all COPUS codes), and two that compare the structures of the classroom observation networks (one measure for instructor networks and one measure for student networks, including all COPUS codes). We used cosine similarity as our similarity metric, as in Ref.~\cite{Sundstrom2025}. Cosine similarity measures the similarity between a pair of vectors, with the cosine of the angle between the vectors being a measure of their effective similarity. Cosine similarity values range from --1 to 1, with values closer to --1 indicating dissimilar vectors, 0 indicating orthogonal vectors, and 1 indicating identical vectors. Mathematically, 
\begin{equation*}
\text{cosine similarity} = 
\frac{\mathbf{v}_1 \cdot \mathbf{v}_2}
{\|\mathbf{v}_1\|\, \|\mathbf{v}_2\|}
\end{equation*}
for two vectors $\mathbf{v}_1$ and $\mathbf{v}_2$.

To measure fidelity based on fractions of class time spent on each COPUS code, the two vectors were defined as the set of these fractions for the high-fidelity and broader implementation being compared. Higher values of cosine similarity indicate that the two instructors had similar distributions of class time spent on the COPUS codes.

To measure fidelity based on classroom observation networks (i.e., capturing the transitions between COPUS codes), the two vectors were defined as the set of network edge weights for the high-fidelity and broader implementations being compared.  Higher values of cosine similarity indicate that the two instructors had similar patterns of temporal sequences of COPUS codes (e.g., spending long amounts of time on certain activities versus cycling through different activities frequently).

\subsubsection{Relating fidelity to student conceptual learning}

For each of the 14 broader implementations with concept inventory data (Table~\ref{observations}), we calculated an effect size using Hedges' \textit{g} to measure student conceptual learning gains. Hedges' \textit{g} is the standardized difference between students' mean pre- and post- concept inventory scores, only including students with matched responses. Hedges'\textit{g} is identical to Cohen's \textit{d}, but includes a correction factor to account for small sample sizes (i.e., small-enrollment physics courses)~\cite{turner2006calculating}. Hedges' \textit{g} has been shown to be more suitable for comparing effect sizes across courses that include both large and small sample sizes~\cite{turner2006calculating}, as we have in this study (Table~\ref{average-classsize} in the Appendix).

To address our third research question, we descriptively examined the relationship between fidelity of implementation (i.e., the four cosine similarity metrics described above) and effect sizes using scatter plots. 

\section{Results}

Below, we present our findings by research question.

\subsection{Fidelity of implementation: Critical components}

Across all three active learning methods, the broad set of instructors implement the critical components to a similar extent as the high-fidelity implementers. In the subsections below, we describe more specific findings for each method.


\begin{figure*}[t]
\includegraphics[width=0.9\textwidth]{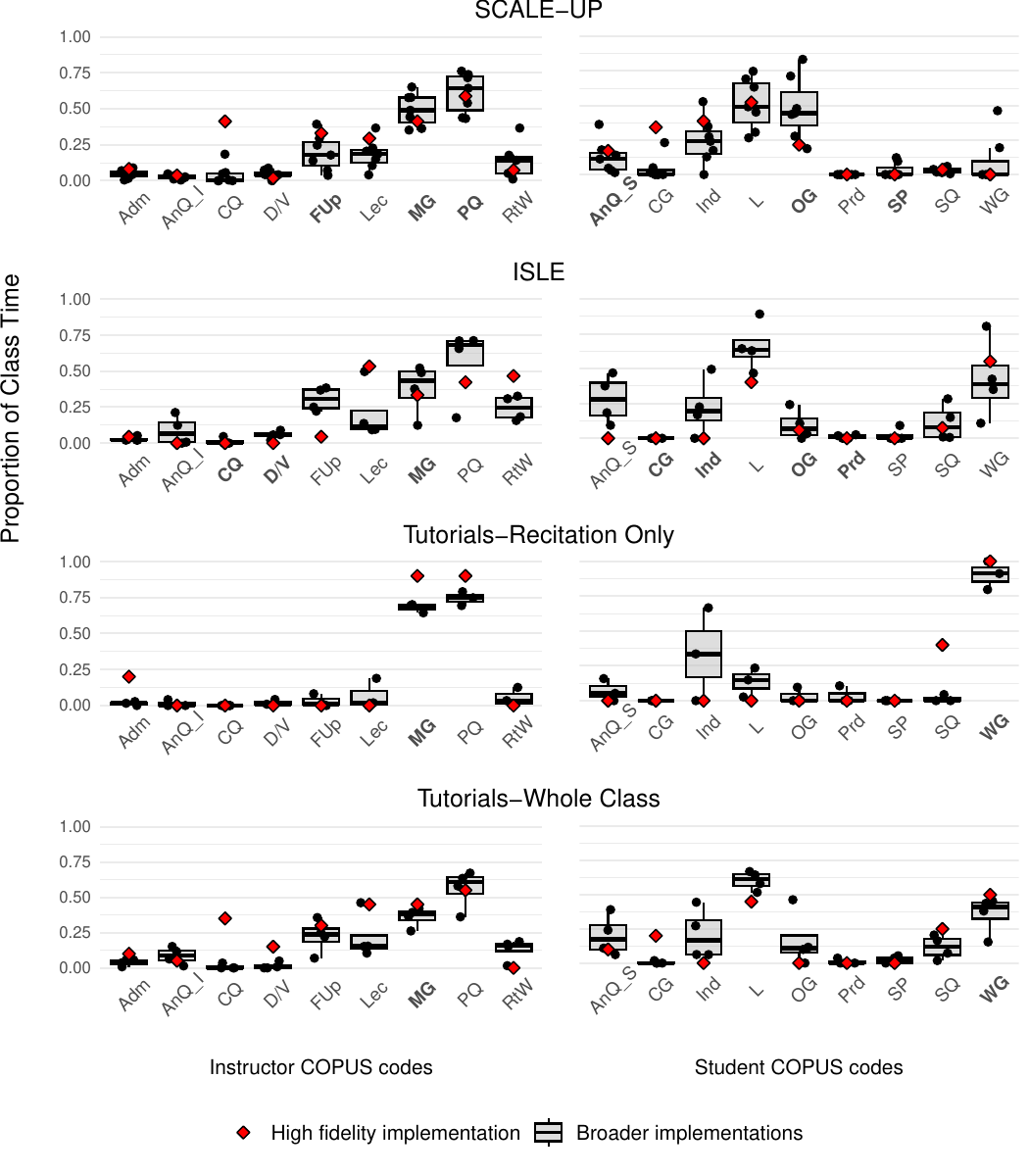}
    \caption{The proportion of two-minute time intervals spent on each COPUS code in high-fidelity and broader implementations of each active learning method. The COPUS codes corresponding to critical components are bolded (Table~\ref{criticalcomponents-table}). The gray boxplots indicate interquartile range, with the bold lines representing the medians and whiskers denoting 1.5 times the interquartile range. }\label{proportion-criticalcomp}
\end{figure*}

\subsubsection{SCALE-UP}

Between the high-fidelity and broader implementations of SCALE-UP, instructors spend similar fractions of class time on all critical components: for the bolded codes, the red diamonds representing the high-fidelity implementations fall within the gray distributions representing the broader implementations in Fig.~\ref{proportion-criticalcomp}. One possible exception to this pattern is student presentations to the whole class (SP), as the high-fidelity implementation of SCALE-UP did not implement this component at all. It is possible that in this course, students presented their solutions or ideas in the form of answers to instructor questions (red diamond is on the upper end of the gray distribution for AnQ\_S in Fig.~\ref{proportion-criticalcomp}).  

Other subtle differences include the broader implementations of SCALE-UP spending less class time on instructor follow-up to activities and more time on students doing group work than the high-fidelity implementation (red diamond is on the upper end of the gray distribution for FUp and the lower end of the gray distribution for OG in Fig.~\ref{proportion-criticalcomp}). The latter may be explained by the prevalence of clicker questions in the high-fidelity implementation (red diamond is much higher than the gray distribution for CQ and CG in Fig.~\ref{proportion-criticalcomp}), indicating that different modes of group work were used in that implementation. 


\begin{figure*}[t]
    \includegraphics[width=1\textwidth]{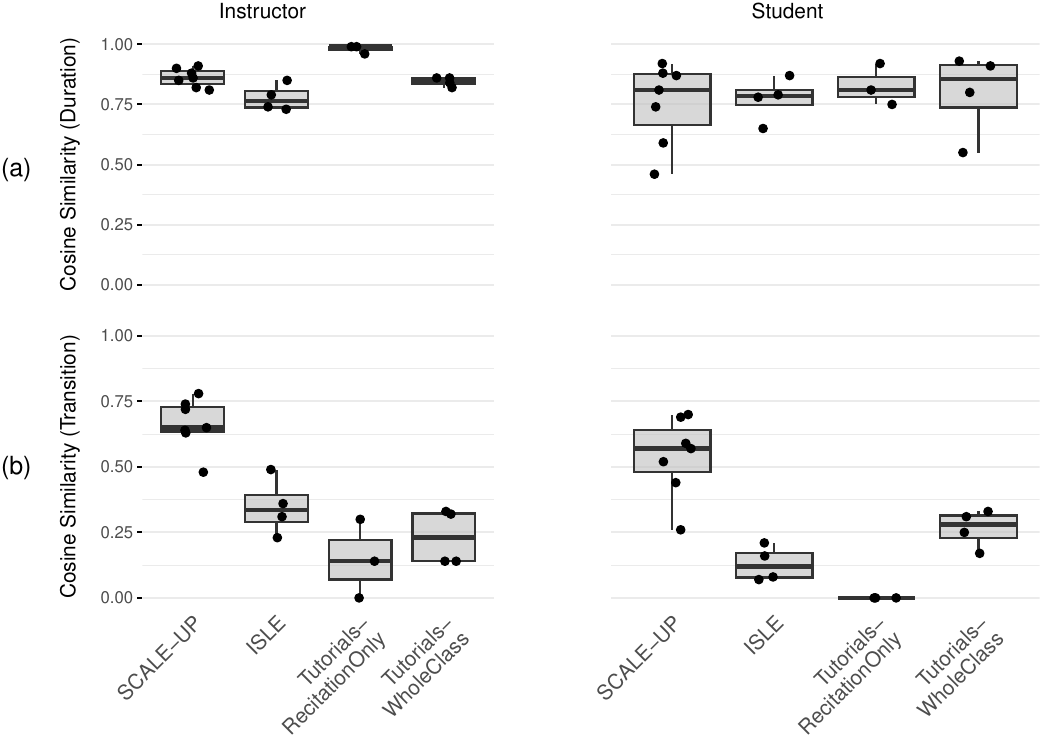}
    \caption{Cosine similarity values comparing (a) the proportion of class time spent on all COPUS activities and (b) patterns of transitions between all COPUS activities (i.e., using the edge weights of classroom observation networks) in broader implementations to high-fidelity implementations of each active learning method. The gray boxplots indicate interquartile range, with the bold lines representing the medians and whiskers denoting 1.5 times the interquartile range.}\label{cosine-sim}
\end{figure*}



\subsubsection{ISLE}

With the exception of the instructor moving and guiding student group work, most of the critical components of ISLE have low prevalence in both the high-fidelity and broader implementations (low values for red diamonds and black dots for all bolded codes except MG in Fig.~\ref{proportion-criticalcomp}). Instead, there is a relatively high prevalence of students completing worksheets in groups in both the high-fidelity and broader implementations, with a slightly higher prevalence of worksheets in the high-fidelity implementation than the broader implementations (red diamond is on the upper end of the gray distribution for WG in Fig.~\ref{proportion-criticalcomp}). As mentioned earlier, the method developers designed worksheets for use in ISLE classrooms~\cite{ISLEweb}. These worksheets include prompts that elicit several of the critical components (e.g., students making predictions about demonstrations). This suggests that both the high-fidelity and broader implementations likely used these ISLE-specific worksheets or operationalized the critical components through their own worksheets.

Another notable difference is that broader implementations of ISLE implemented individual student work more than the high-fidelity implementation (red diamond is on the lower end of the gray distribution for Ind in Fig.~\ref{proportion-criticalcomp}). Individual work is often coded alongside working on clicker questions in groups if there are some students working alone, suggesting the high-fidelity implementation engaged all students in small group discussions when completing these types of questions.


\subsubsection{Tutorials}

In both the whole-class and recitation-only versions of Tutorials, the high-fidelity and broader implementations have high and comparable prevalence of the two critical components: the instructor moving and guiding group work and the students completing worksheets in groups (similarly high values of red diamonds and black dots for MG and WG codes in Fig.~\ref{proportion-criticalcomp} for each version). In the recitation-only version of Tutorials, the instructor moved and guided the group worksheets in the high-fidelity implementation more than in the broader implementations (red diamond is above the gray distribution for MG in Fig.~\ref{proportion-criticalcomp}). In general, worksheets are used less frequently in the whole-class version of Tutorials than the recitation-only version, which aligns with the goals of each of these versions of the method (i.e., Tutorials supplementing lecture and other activities in the whole-class version versus Tutorials being the main activity of recitation section in the recitation-only version; higher values of red diamonds and gray distributions for WG in the recitation-only than whole-class version in Fig.~\ref{proportion-criticalcomp}).



\subsection{Fidelity of implementation: Full COPUS observations}

We use cosine similarity to compare both the proportions of class time spent on each COPUS code (including all COPUS codes, not only those related to critical components, because some critical components may be operationalized through different mediums such as worksheets) and the classroom observation networks (which capture temporal transitions between all COPUS codes) between high-fidelity and broader implementations of each active learning method (Fig.~\ref{cosine-sim}).


\subsubsection{Proportion of time spent on classroom activities}

With regard to the proportion of class time spent on each COPUS code, there is high fidelity for all broader implementations of all three methods (cosine similarity values close to one for both instructor and student codes in each active learning method in Fig.~\ref{cosine-sim}a). Only considering the student codes, there is comparable fidelity across all active learning methods (the gray distributions overlap for student codes in Fig.~\ref{cosine-sim}a). Only considering the instructor codes, there is variation in fidelity across methods: the recitation-only version of Tutorials has the highest fidelity, followed by SCALE-UP, the whole-class version of Tutorials, and then ISLE. This pattern of fidelity is largely consistent with that found in our descriptive comparison of the prevalence of critical components presented in the previous section.

Additionally, there is more variation in fidelity within each active learning method for student codes than instructor codes (for each method, the gray distribution for student codes spans a wider range than the gray distribution for the instructor codes in Fig.~\ref{cosine-sim}a). This suggests that the same instructor code may correspond to different student codes in different implementations. For example, if an instructor poses a question or activity to the whole class, PQ, this student activity may be in the form of worksheets, WG, or other group work, OG.

\begin{figure}[t]
    \includegraphics[width=0.55\textwidth]{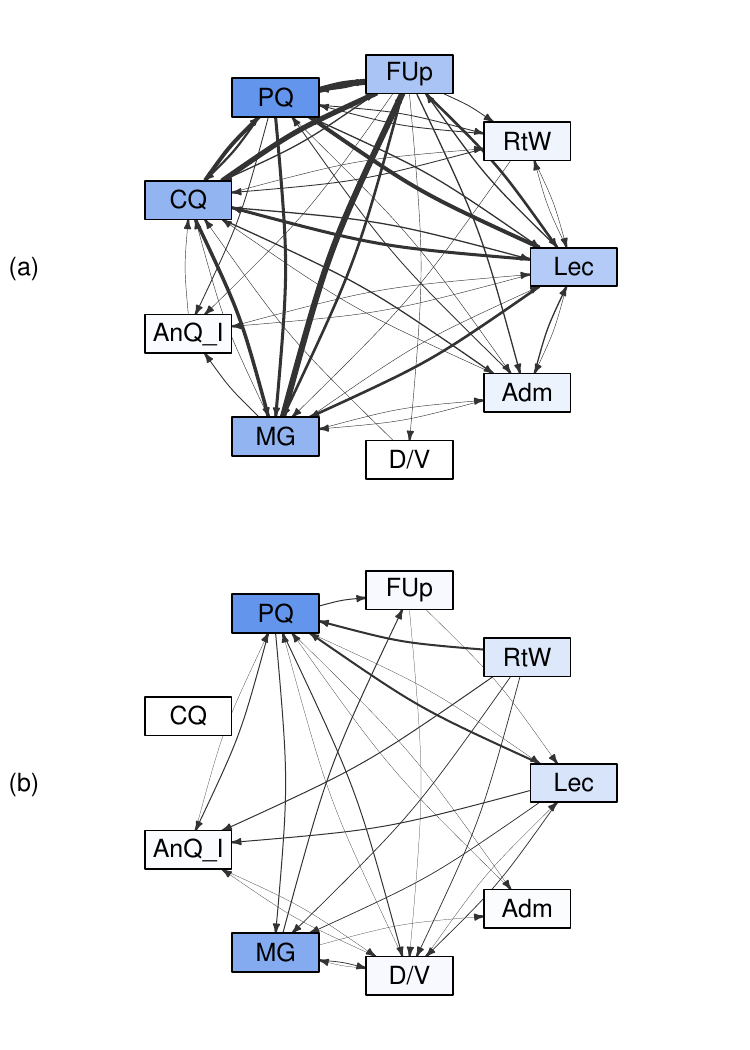}
    \caption{Example classroom observation networks, only considering instructor codes, for (a) the high-fidelity implementation of SCALE-UP and (b) one broader implementation of SCALE-UP with high cosine similarity values for duration fidelity and low cosine similarity values for transition fidelity. Node color represents the fraction of observed two-minute time intervals spent on each code, with darker shades indicating larger fractions. Edges point from an initial code to the code that occurs in the following two-minute time interval of the COPUS observation. Edge width indicates the number of transitions occurring between those two codes normalized by the total number of two-minute time intervals observed. Networks for all courses in this study are available at Ref.~\cite{github2025}.}\label{network-example}
\end{figure}

\begin{figure*}[t]
    \includegraphics[width=1\textwidth]{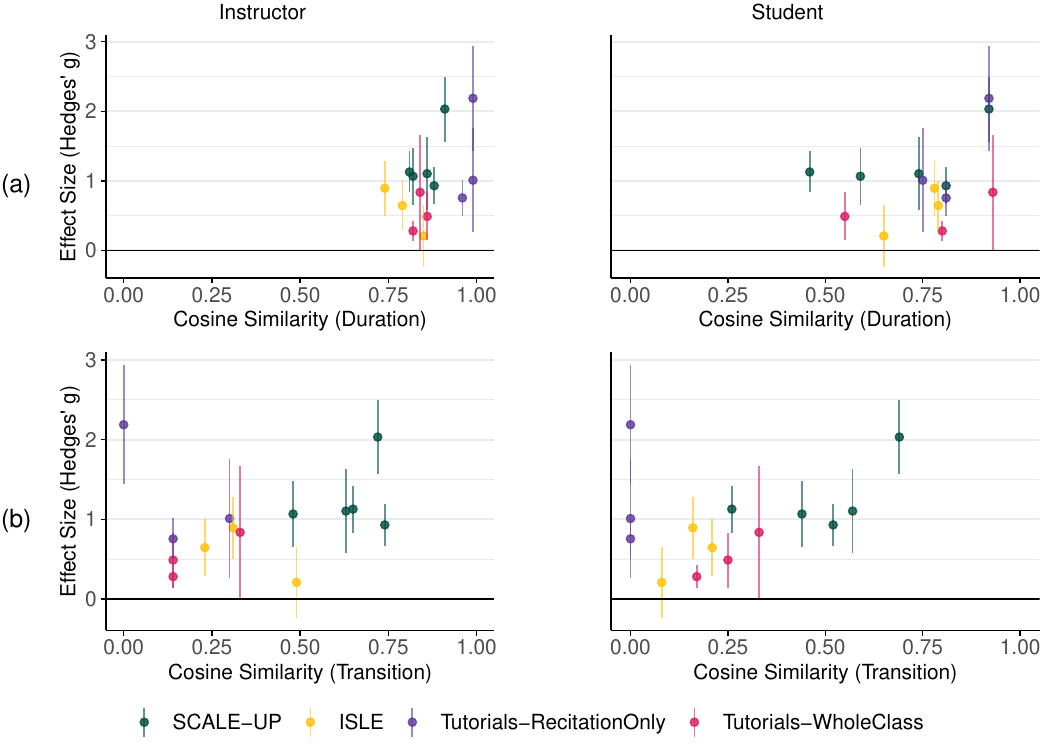}
    \caption{Effect sizes of student concept inventory scores versus (a) duration fidelity, measured using the proportion of class time spent on all COPUS codes, and (b) transition fidelity, measured using classroom observation networks (Fig.~\ref{cosine-sim}). Dots indicate Hedges' \textit{g} values and error bars indicate 95\% confidence intervals.}\label{learning-gains}
\end{figure*}

\subsubsection{Transitions between classroom activities}

With regard to classroom observation networks, which capture the chronological patterns of activities, we observe lower cosine similarity values across all active learning methods than the fidelity measured with regard to proportions of class time spent on each activity (for each method, cosine similarity values are lower in Fig.~\ref{cosine-sim}b than Fig.~\ref{cosine-sim}a). This indicates that even when instructors spend similar amounts of time on certain classroom activities, those activities are executed in very different ways (e.g., short, discontinuous sequences of activities versus long, continuous activities). For example, Fig.~\ref{network-example} shows the classroom observation networks for the high-fidelity implementation of SCALE-UP and one broader implementation of SCALE-UP with high cosine similarity values for duration fidelity and low cosine similarity values for transition fidelity (only for the instructor codes). We see that the node colors, which indicate the proportion of time spent on the codes, are mostly comparable between the two networks (with the exceptions of clicker questions, CQ, and following up, FUp). The broad implementation (Fig.~\ref{network-example}b), however, generally has thinner edge weights than the high-fidelity implementation (Fig.~\ref{network-example}a), indicating that the broad implementation includes extended time periods of activities and the high-fidelity implementation cycles through shorter versions of similar activities.

Unlike fidelity based on duration of class activities, there is substantial variation in fidelity based on transitions between class activities across the active learning methods. SCALE-UP seems to be implemented with higher fidelity than the other active learning methods when we consider the chronological patterns of activities (gray distributions are higher for SCALE-UP than the other methods for both instructor and student codes in Fig.~\ref{cosine-sim}b). We note that some cosine similarity values are equal to zero for the recitation-only version of Tutorials because those classroom observation networks are very sparse, as the class sessions often involve students doing worksheets for the whole class with very few transitions to or from other activities.


\subsection{Relating fidelity of implementation to student conceptual learning}

There is no clear relationship between fidelity of implementation of the observed active learning methods, as measured by either duration of class time spent on each COPUS code or the transitions between COPUS codes captured in the classroom observation networks, and student conceptual learning (Fig.~\ref{learning-gains}).
For fidelity measured using duration of activities, all cosine similarity values are fairly high, so we may not have enough variation of fidelity in our data to determine if any relationship exists (Fig.~\ref{learning-gains}a). For fidelity measured using transitions between activities, we have more variation in cosine similarity values, but there is no clear upward, downward, or flat trend in the scatterplot (Fig.~\ref{learning-gains}b). There is some preliminary evidence that SCALE-UP has both higher transition fidelity and higher student learning gains than the other methods, but we cannot make any strong conclusions about particular active learning methods based on our small sample size (green dots are clustered in the top right of Fig.~\ref{learning-gains}b for both instructor and student codes).

\section{Discussion}

In this study, we measured the fidelity of implementation of 18 introductory physics instructors using one of three named active learning methods: SCALE-UP, ISLE, and Tutorials. We did so using direct classroom observations and the COPUS, which is likely more accurate than self-reported instructor practices collected through survey or interviews~\cite{borrego2013fidelity,dancy2016faculty}. We analyzed COPUS data in two ways: first, by measuring the extent to which the critical components of each method were implemented by the 18 instructors as compared to high-fidelity implementers, and second, by comparing the full COPUS observations of these broader implementations to those of high-fidelity implementations.  We also expand upon existing studies of fidelity in undergraduate science courses by relating fidelity to student conceptual learning.


\subsection{Fidelity of implementation of named active learning methods}




We found that broader implementations of active learning mostly use the critical components of the methods. Previous studies have measured fidelity solely based on whether or not each critical components is present in the implementation; by this definition, most of the instructors in our study implemented the method with fidelity~\cite{borrego2013fidelity, dancy2016faculty}. However, we also measured the fractions of class time spent on each critical component and the fractions of class time spent on all COPUS codes. These fine-grained measurements again highlighted that broader implementations used the critical components, and other classroom activities, to similar extents as the high-fidelity implementations. These results may suggest that active learning methods are generally being implemented as intended by the developers; however, this pattern could be due to the nature of our study sample. Most of the instructors in the this study are either currently conducting physics education research, previously conducted physics education research (i.e., during graduate school), or are high consumers of physics education research. Prior work has shown that instructors trained in science education research may teach differently, and have more positive impacts on student outcomes, than a random sample of instructors~\cite{andrews2011active}. We encourage future work to measure the fidelity with which a more representative sample of college physics instructors implement named active learning methods to extend our findings.

We also observed that in some active learning methods, instructors spend a relatively high proportion of time on non-critical component codes, for example posing questions (PQ) in ISLE and the recitation-only version of Tutorials. Although these codes do not directly reflect critical components of the methods, they are central to active learning instruction and facilitate the use of critical components (e.g., moving and guiding groupwork, MG, and posing questions are often coded concurrently). 

At the same time, we observed variation in the ways in which broader implementers used these critical components and class activities (i.e., based on classroom observation networks that capture the chronological transitions between activities~\cite{Sundstrom2025}). That is, while broader implementation instructors spend similar proportions of time on activities as the high-fidelity implementations, they differ in how they distribute these activities across class time. One of the SCALE-UP instructors, for example, used a few long time periods to implement groupwork activities. The high-fidelity implementation of SCALE-UP, on the other hand, implemented the same types of activities in many short time periods. Future research should continue to characterize these different instructional styles and determine the method developers' intentions for specific ways of implementing the activities (i.e., to validate the fidelity of transitions between activities in addition to the critical components). 

Interestingly, when considering these chronological patterns of COPUS codes, broader implementations of SCALE-UP had noticeably higher fidelity than the broader implementations of ISLE and Tutorials. This pattern is surprising because both ISLE and Tutorials have more specific curricula (e.g., with carefully designed worksheets) than SCALE-UP, which places stronger emphasis on the physical classroom layout~\cite{physport2025}. One hypothesis is that the structural features of SCALE-UP classrooms inherently constrain instructional choices, allowing for consistent instructor enactment of critical components even without highly prescriptive materials. Indeed, the broader implementations of ISLE and Tutorials courses took place in a wide variety of classrooms, possibly leading to a wide range in instructional practices. Another explanation is that instructors who adopt SCALE-UP receive more focused training or institutional support related to its implementation, which could lead to higher fidelity. We recommend for future studies to examine these possibilities for variation in fidelity across the active learning methods.



\subsection{Fidelity of implementation and student conceptual learning}

Our analysis revealed no clear relationship between fidelity of implementation and conceptual learning gains, as measured using concept inventories. We did, however, identify preliminary evidence that SCALE-UP courses tend to be implemented with higher fidelity (particularly as measured using the classroom observation transition networks) than the other methods. This may (at least partly) explain results from another study of these data documenting significantly higher student learning gains in SCALE-UP than ISLE~\cite{sundstrom2025relativebenefits}. Perhaps fidelity of implementation is necessary, but not sufficient for learning gains (i.e., other factors such as class size, student preparation, and homework assignments may also matter)~\cite{Sundstrom2025}. This proposition aligns with the findings of prior work related to ``reinvention" of research-based instructional strategies, where the researchers found a correlation between fidelity and outcomes of social innovation programs, and that local additions to the critical components further enhance impacts to outcomes~\cite{blakely2002fidelity}.

Further research is needed to better understand the role of fidelity in the effectiveness of active learning methods to student outcomes in undergraduate physics (and science) courses, including a dataset with more instructors and more student outcomes (i.e., attitudes, identity, and experimental skills) than the study presented here. If there is a relationship between fidelity and student outcomes, this would indicate that preserving the  critical components \textit{and} the specific ways they are operationalized is necessary for improving student outcomes. If there is no strong relationship between fidelity of implementation and student outcomes, this would imply that instructors can flexibly choose the style in which they implement the critical components without sacrificing effectiveness. 


\subsection{Limitations and future work}

While this study offers valuable insights to the fidelity of implementation of named active learning methods in introductory physics and astronomy, there are several limitations that may be addressed in future research. First, we used extensive literature review and negotiated discussions among the research team to identify the critical components of each active learning method and their mapping to COPUS codes. Future research that validates these sets of critical components, their corresponding COPUS codes, the fraction of class time instructors should spend on these codes, and the intended flow of or transitions between activities (e.g., through developer and/or expert interviews) will provide further context for our claims. 


Second, we have assumed that high-fidelity implementation data from the first iteration of the project is indeed high fidelity. This is a plausible assumption because these observations were largely conducted at the development sites of the methods in our study; however, it is possible that over time (i.e., between the time of development and the implementation we analyzed), fidelity of implementation might have changed~\cite{kotter2007leading,foote2016enabling}. Relatedly, the previous researchers did not conduct some of the high-fidelity observations for full class sessions (i.e., they observed 20 min segments of some class sessions), though they noted that the observed subset of the full class time was representative~\cite{commeford2021characterizing, commeford2022characterizing}. Future studies should conduct multiple, full-length classroom observations of high-fidelity implementations and/or aim to observe more than one high-fidelity implementation of each named active learning method to validate our findings~\cite{lund2015best,weir2019small}. 

Finally, in terms of the COPUS coding, the high-fidelity and broader implementation data were collected and coded several years apart, so we could not establish inter-rater reliability across the coders involved in each iteration of the project. This may have led to subtle differences in the ways the codes were applied to the observations in each dataset; however, the COPUS was designed for consistent application across many observers with minimal training, reducing the likelihood that small variations in coding meaningfully affected the overall patterns in our analysis~\cite{smith2013classroom}. Additionally, the COPUS may not capture the full range and nuance of classroom activities (e.g., when student predictions are embedded in group worksheets and when the same instructor activity coincides with multiple different student activities), so our analysis may have overlooked some aspects of the courses that are related to fidelity. We recommend for future work to examine possible improvements to the COPUS or other observation protocols and/or supplement direct classroom observations with other course information, such as instructor interviews or course materials, to more accurately measure fidelity.


\section{Conclusions}

We have showcased a novel approach to measuring fidelity that we hope other researchers will use to better understand the gray area often associated with fidelity of implementation of active learning strategies. In our study of 18 introductory physics instructors using SCALE-UP, ISLE, or Tutorials, we found generally high fidelity of implementation and no clear relationship between fidelity and student conceptual learning. Future studies should validate and improve our approach and measure fidelity and its impacts across a more diverse set of instructors, including those outside of science education research, those in other scientific disciplines, and those outside of the United States.  Such studies will help to narrow down the conditions under which active learning methods are the most effective for student outcomes, which will inform best practices for undergraduate science education.

\acknowledgments{We thank the instructors and students who participated in our study. We also thank David Esparza, Ian Olivant, and Bruna Schons Ribeiro for their feedback on this manuscript. This work is supported by the National Science Foundation Grant Nos. 2111128 and 2139899, and the Cotswold Foundation Postdoctoral Fellowship at Drexel University.
}

\bibliography{bibfile} 

\clearpage

\section*{APPENDIX}

\subsection{Concept inventories}
Table~\ref{conceptinventory-breakdown} summarizes the concept inventories used to measure student learning by active learning method.

\subsection{Class sizes}
Table~\ref{average-classsize} summarizes the average number of enrolled students per course in each active learning method.

\begin{table*}[h]
  \caption{Number of broader implementation courses using each concept inventory by active learning method. Concept inventory data were not collected for the high-fidelity implementations.\label{conceptinventory-breakdown}}
  \begin{ruledtabular}
    \begin{tabular}{lccc}
 Method & SCALE-UP & ISLE & Tutorials \\
   \hline
Force Concept Inventory~\cite{hestenes1992force}&4 &2 &3\\
Force and Motion Concept Evaluation~\cite{ramlo2008validity}&0&1&1\\
Energy and Momentum Conceptual Survey~\cite{singhemcs}&1&0&0\\
Energy Concept Assessment~\cite{eca}&0&0&1\\
Light and Spectroscopy Concept Inventory~\cite{bardar2007development}&0&0&1\\
    \end{tabular}
  \end{ruledtabular}
\end{table*}

\begin{table}[h]
 \caption{Class sizes (i.e., number of enrolled students) by active learning method. For the high-fidelity implementation of Tutorials, the recitation-only and whole-class observations come from the same course, so the number of enrolled students is the same. For broader implementations, the values indicate means, with standard deviations in parentheses. \label{average-classsize}}
 \begin{ruledtabular}
    \begin{tabular}{lcc}
    Method & High-fidelity & Broader\\
    & implementation & implementations\\
    \hline
        SCALE-UP & 71 & 64.1 (38.8) \\
        ISLE & 28 & 24.0 (9.8)\\
        Tutorials (Recitation-only) & 171 & 30.7 (19.0)\\
        Tutorials (Whole-class) & 171 & 38.3 (22.4)\\
    \end{tabular}
 \end{ruledtabular}
\end{table}
\end{document}